\documentclass[conference, 10pt]{IEEEtran}
\ifCLASSINFOpdf
\else
\fi
\usepackage[cmex10]{amsmath}
\usepackage{amsfonts}
\usepackage{multirow}
\makeatletter
\newcommand{\xRightarrow}[2][]{\ext@arrow 0359\Rightarrowfill@{#1}{#2}}
\makeatother

\usepackage{algorithm}
\usepackage{algorithmic}

\usepackage{tikz}
\usepackage{textcomp}
\usepackage{hyperref}
\usepackage{lipsum}

\newcommand\copyrighttext{%
  \footnotesize \textcopyright 2016 IEEE. Personal use of this material is permitted.
  Permission from IEEE must be obtained for all other uses, in any current or future
  media, including reprinting/republishing this material for advertising or promotional
  purposes, creating new collective works, for resale or redistribution to servers or
  lists, or reuse of any copyrighted component of this work in other works.
}
\newcommand\copyrightnotice{%
\begin{tikzpicture}[remember picture,overlay]
\node[anchor=south,yshift=10pt] at (current page.south) {\fbox{\parbox{\dimexpr\textwidth-\fboxsep-\fboxrule\relax}{\copyrighttext}}};
\end{tikzpicture}%
}

\usepackage[tight,footnotesize]{subfigure}

\usepackage{graphicx}
\usepackage{color}
\usepackage{placeins}
\usepackage{float}
\usepackage{hyperref}
\usepackage{tabularx,colortbl}

\begin{document}
%
\title{Direct Domain Decomposition Method (D$^3$M) for Finite Element Electromagnetic Computations}

\author{\IEEEauthorblockN{Javad Moshfegh and Marinos N. Vouvakis}
\IEEEauthorblockA{Department of Electrical and Computer Engineering, University of Massachusetts, Amherst, MA, USA}}

\maketitle
\copyrightnotice

\begin{abstract}
An exact arithmetic, memory efficient direct solution method for finite element method (FEM) computations is outlined. Unlike conventional black-box or low-rank direct solvers that are opaque to the underlying physical problem, the proposed method leverages physical insights at every stage of the development through a new symmetric domain decomposition method (DDM) with one set of Lagrange multipliers. Comparisons with state-of-the-art exact direct solvers on electrically large problems suggest up to 10 times less memory and better run-time complexity while maintaining the same accuracy.
\end{abstract}
\vspace{8pt}
This work has been presented at the 2016 International Conference on Electromagnetics in Advanced Applications. The parallel version of this work has been presented at the 2019 IEEE International Symposium on Antennas and Propagation and USNC-URSI Radio Science Meeting [\ref{ref:moshfegh2019}]
\IEEEpeerreviewmaketitle

\section{Introduction}
Modern computational electromagnetics FEM tools often rely on memory efficient iterative solvers such as Multigrid [\ref{ref:zhu2006}] and DDM [\ref{ref:toselli2005}] that may experience convergence difficulties near resonances or multi-scale problems, and lose efficiency at multiple excitation runs. Contrary, direct solvers such as MUMPS [\ref{ref:amestoy2001}] or PARDISO are reliable but scale unfavorably and are hard to parallelize. Thus, recent trends in direct solvers [\ref{ref:ying2014}] strive to reduce workload by leveraging low-rank approximations at the cost of accuracy and possibly reliability. Yet, these solvers are opaque to important underlying physics, leaving room for further improvements.

To achieve an efficient $exact$ direct solver, one must start from scratch and attempt to leverage deep physical and numerical insights that may require re-formulating the BVP and FEM assembling, in addition to the symbolic and numeric factorization stages. This is critical not only to avoid internal resonances at all intermediate factorization separators, but also to produce numerically efficient matrix structures i.e. reduced size, block-wise sparse symmetric matrices.

This work achieves all these via a direct DDM (D$^3$M) framework. A set of auxiliary variables is used to cast a decomposed BVP that, after an initial reduction/elimination step, leads to an auxiliary blocked matrix that is suitable for factorization. To attain maximal performance, this matrix is factored with a special blocked LDL$^T$ method with restricted Bunch-Kaufman pivoting [\ref{ref:bunch1977}].

The accuracy and performance of the proposed D$^3$M has been verified and tested in 3D scattering problems by perfect electric conductor (PEC) plates and dielectric spheres of progressively larger electrical sizes. The proposed D$^3$M solver requires $3\!\times\! - \!10\times$ less memory than MUMPS mainly due to the choice of structured separators and the absence of delayed pivots attributed to the interior resonance free formulation. An initial serial implementation of D$^3$M was up to two times slower than MUMPS for small problems but becomes competitive on problems larger than one million unknowns.
\section{Theory}
\begin{figure}[h]
\begin{center}
\noindent
  \includegraphics[width=2.5in]{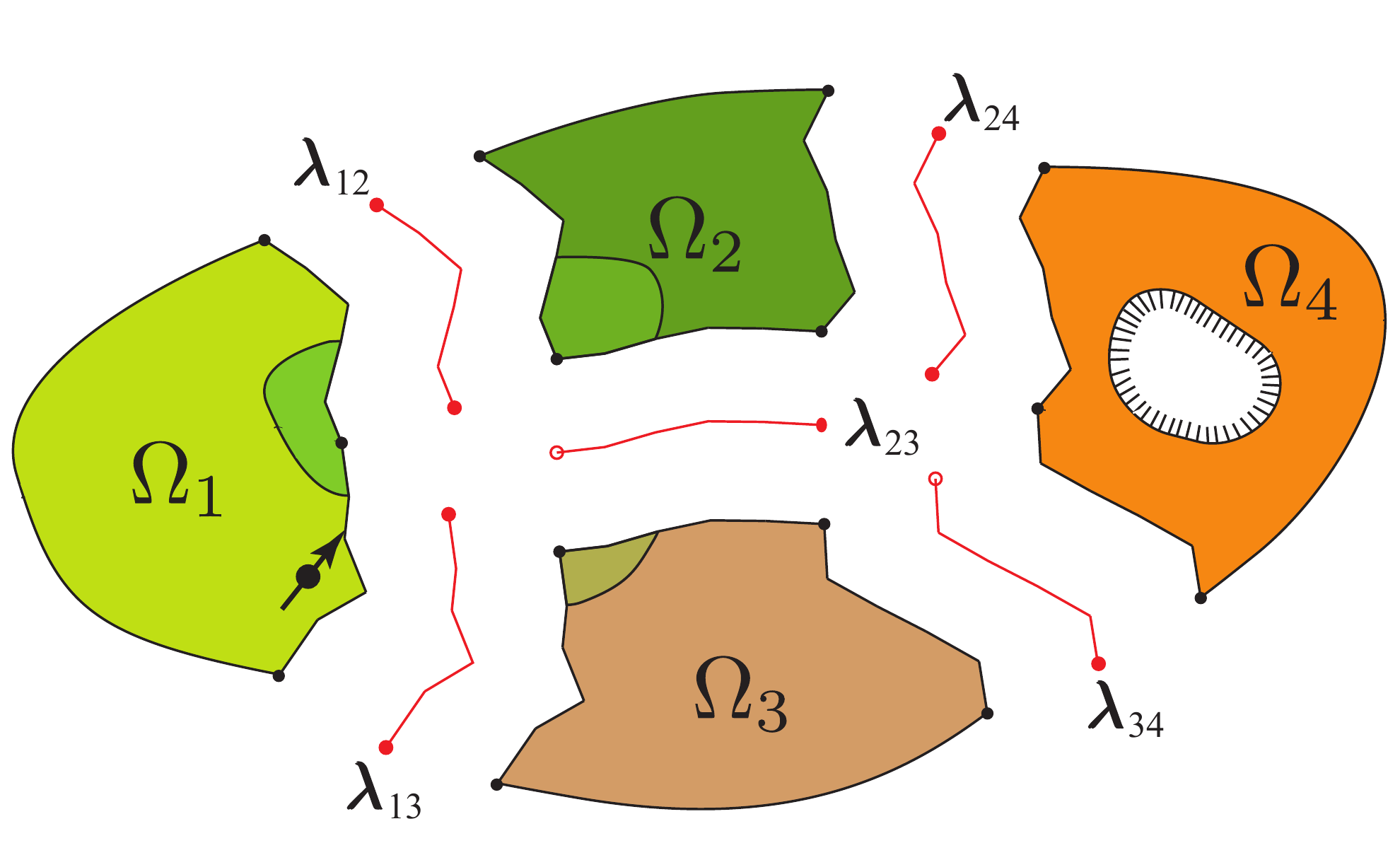}
\vspace{-10pt}
  \caption{2D decomposition problem with four domains.}\label{fig:2Ddecomp}
\end{center}
\end{figure}

Consider a computational domain $\Omega$ decomposed into $N$ non-overlapping subdomain $\Omega_i$. For example, a decomposed problem with four domains is shown in Fig. \ref{fig:2Ddecomp}. The decomposed BVP reads, $find \left({\bf E}_{*}, \pmb{\lambda} \right) \in \left\{{\bf V}_{*}, \pmb{\Lambda}\right\} such \mbox{ } that$:
\begin{equation} \nonumber
\left\{ 
\begin{array}{ll}
\nabla\times\frac{1}{\mu_r}\nabla\times {\bf E}_{*} -k^2\epsilon_r  {\bf E}_{*} = -j
k\eta{\bf J}^{imp}, \hspace{15pt} \mbox{in } \, \Omega = \bigcup \Omega_i\\
\gamma_{\times}\!\!\left(\! \nabla\!\!\times\!\!{\bf E} \right) \!- \!jk \gamma_{t}\left( {\bf E}\! \right) \! =\!\gamma_{\times}\!\left( \nabla\!\times\!{\bf E}^{inc} \right) \!- \!jk \gamma_{t}\left( {\bf E}^{inc}\! \right)\!, \hspace{1pt} \! \mbox{on } \, \partial\Omega \\
{\bf R}^{(i)}_{ij}\left[\mathbf{j} - \alpha \mathbf{e}\right]-{\bf R}^{(\mathcal{N}(i))}_{ij}\left[-\mathbf{j} - \alpha \mathbf{e}\right] = {\bf 0}, \! \hspace{15pt}\! \mbox{on } \, {\cal S}, \forall i \,\&\, \{ij\} \\
{\bf R}^{(i)}_{ij}{\bf e}-{\bf R}^{(\mathcal{N}(i))}_{ij}{\bf e} = {\bf 0},  \hspace{80pt}\! \mbox{on } \, {\cal S}, \forall i\,\&\, \{ij\}
\end{array} 
\right.
\end{equation}
where $\mathbf{e} = \gamma_t(\mathbf{E}|_\mathcal{S})$, and $\mathbf{j} = \gamma_\times(\nabla \!\times\! \mathbf{E}|_\mathcal{S})$. ${\bf R}^{(i)}_{ij}$ is the restriction operator from domain $i$ to interface $ij$, and $\mathcal{N}(i)$ denotes the neighbor of domain $i$. \\
\vspace{-20pt}
\vspace{-10pt}
After transforming to $\pmb{\lambda}=\bf{j} \pm \mathbf{e}$, casting the variational problem and expanding trial and testing  function spaces gives 

\vspace{-10pt}
\begin{equation}
		\!\!\left[\!\! \begin{array}{cc} \!\!\mathbf{A}\!\! & \!\!\mathbf{D}\!\! \\
		\!\!\mathbf{D}^{T}\!\! & \!\!\mathbf{0}\!\! \\
		\end{array}
		\!\!\right]\!\!\!\left[\!\!
		\begin{array}{c}
		\!\mathbf{E}\!\!\!\\
		\!\pmb{\lambda}\!\!\!\\
		\end{array}
		\!\!\right]\!\!=\!\!\left[\!\!\!
		\begin{array}{c}
		{\bf f}\\
		{\bf 0}\\
		\end{array}
		\!\!\!\right]
		\!\xRightarrow{Elim.\mathbf{E}}\!
		\!\underbrace{\left(\sum\limits_{i=1}^{N_I} \mathbf{D}^T_i (\mathbf{A}_i)^{-1} \mathbf{D}_i\!\right)}_\text{\bf{K}} \! \pmb{\lambda} = \mathbf{g},
\end{equation}
where $\mathbf{A}$ is the diagonal blocked matrix of $\mathbf{A}_i, i = 1,2,...,N$ where $\mathbf{A}_i$ is the FEM-ABC matrix for domain $i$ with loss or gain at the interfaces. $\mathbf{D}$ is a blocked matrix of sparse matrices $\mathbf{D}_i, i = 1,2,...,N_I$ which maps the primal space to LM space. $N_I$ is the number of interfaces.

The reduced matrix ${\bf K}$ is symmetric block-wise sparse but indefinite. Hence LDL$^T$ factorization with symmetric partial pivoting a.k.a Bunch-Kaufmann LDL$^T$ [\ref{ref:bunch1977}] can be used to save memory and CPU time. Since ${\bf K}$ is a block-wise sparse matrix, we have modified the Bunch-Kaufman LDL$\!^T$ factorization to its block restricted partial pivoted form. Each block in ${\bf K}$ corresponds to a super-node of typical order $n > 300$. Therefore, D$^3$M consistently operates at the maximum performance region of Level 3 BLAS. The main steps of the proposed D$^3$M are:

\begin{enumerate}
\item Generate dense domain matrices $\!\mathbf{K}_D^{(d)}\!$,
\item Assemble the block-wise sparse reduced matrix ${\bf K}$,
\item Reorder the clique graph of ${\bf K}$,
\item Symbolic factorize the reordered clique graph,
\item Factorize ${\bf K}$ with restricted B-K pivoting block LDL$\!^T$ factorization (see algorithm 1),
\item Forward/Backward substitute the reduced system ${\bf K}$ for auxiliary unknowns,
\item Recover primal unknowns.
\end{enumerate}

The clique graph of blocked sparse matrix ${\bf K}$ is reordered using METIS (same as MUMPS). Assuming that the clique graph has $l$ levels, the algorithm of block LDL$^T$ factorization is given in Algorithm 1. A multifrontal version of block LDL$^T$ can be used to further speed computations.
\vspace{-5pt}
\begin{algorithm}
\caption{Block LDL$\!^T$ factorization}
\begin{algorithmic}
\FOR {$j = 1 \rightarrow l$}
\STATE Dense LDL$^T$ factorize ${\bf K}_{jj}$
\FOR {$\forall i | {\bf K}_{ij} \neq 0$}
\STATE Dense Triangular solve $\mathbf{L}_{jj} \mathbf{X}_{ij}^T  \!\!=\!\! \mathbf{K}_{ij}^T,$ for $\mathbf{X}_{ij} \!\!=\!\! \mathbf{P}_{ii} \mathbf{L}_{ij} \mathbf{D}_{jj}$
\STATE Compute work variable $\mathbf{W}_{ij} =\mathbf{X}_{ij} \mathbf{D}_{ii}^{-1}$
\FOR {$\forall k | {\bf K}_{ik} \neq 0$}
\STATE Dense update  $\mathbf{K}_{ik}  \leftarrow \mathbf{A}_{ik} \!-\! \mathbf{X}_{ij}\mathbf{W}_{ij}^T$
\ENDFOR
\ENDFOR
\ENDFOR
\end{algorithmic}
\end{algorithm}
\vspace{-5pt}
\section{Numerical Results}
\begin{figure}
	\begin{center}
		{\includegraphics[width=2.5in]{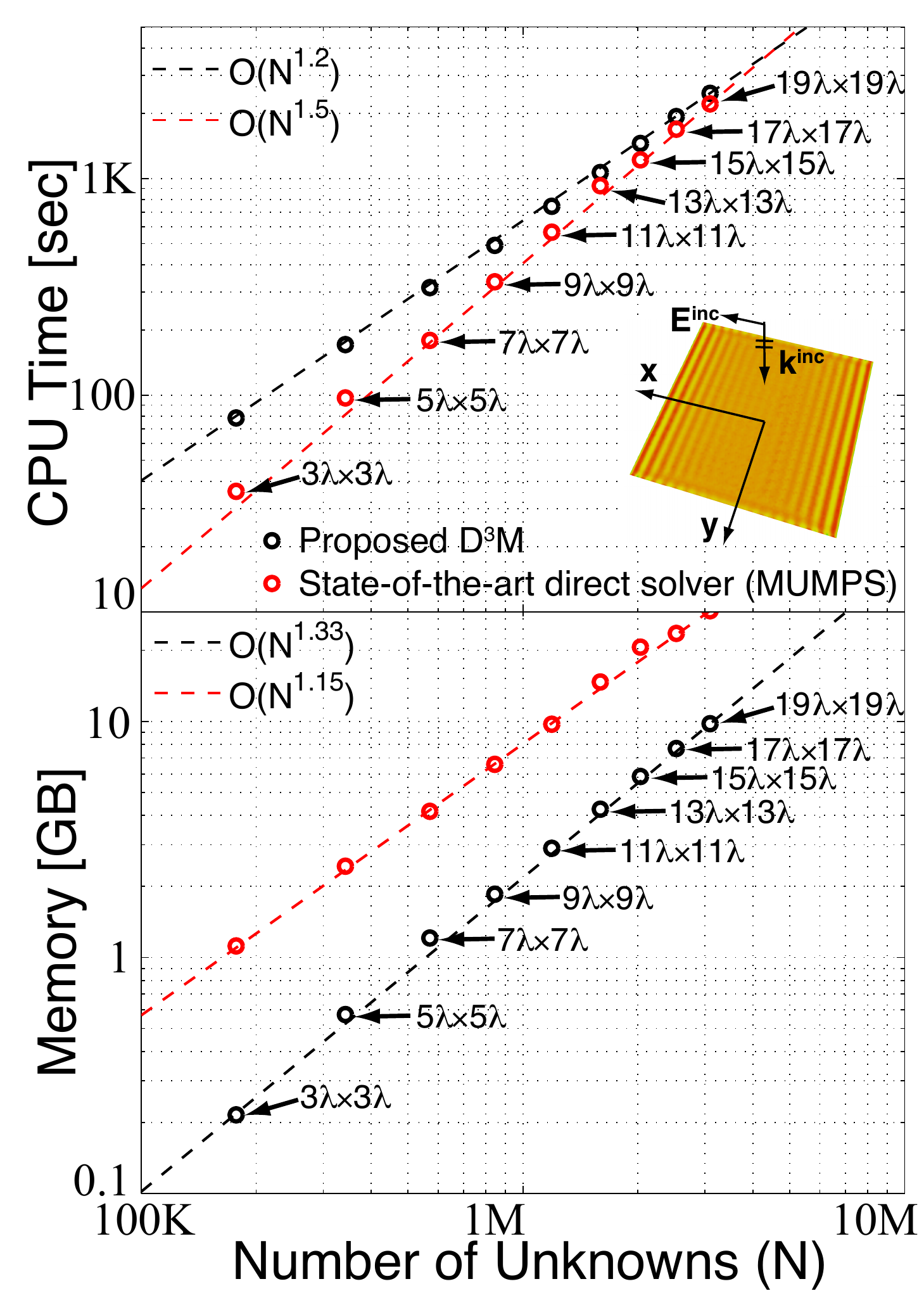}}
		\caption{Computational complexity for progressively larger PEC plates using the proposed D$^3$M and MUMPS: Factorization time (top); Factorization memory (bottom).}\label{fig:pec_plate}
	\end{center}
 \end{figure}

First, the scattering of progressively larger PEC plates (from $3\lambda \!\times\! 3\lambda$ up to $19\lambda \!\times\! 19\lambda$) are considered. Computational complexity of factorization time and memory for these problems using the proposed D$^3$M and MUMPS are shown in Fig. \ref{fig:pec_plate}. It is noted that a 3M unknown problem is solved with only 10 GB of RAM at full double precision accuracy.

\begin{figure}
	\begin{center}
		{\includegraphics[width=2.5in]{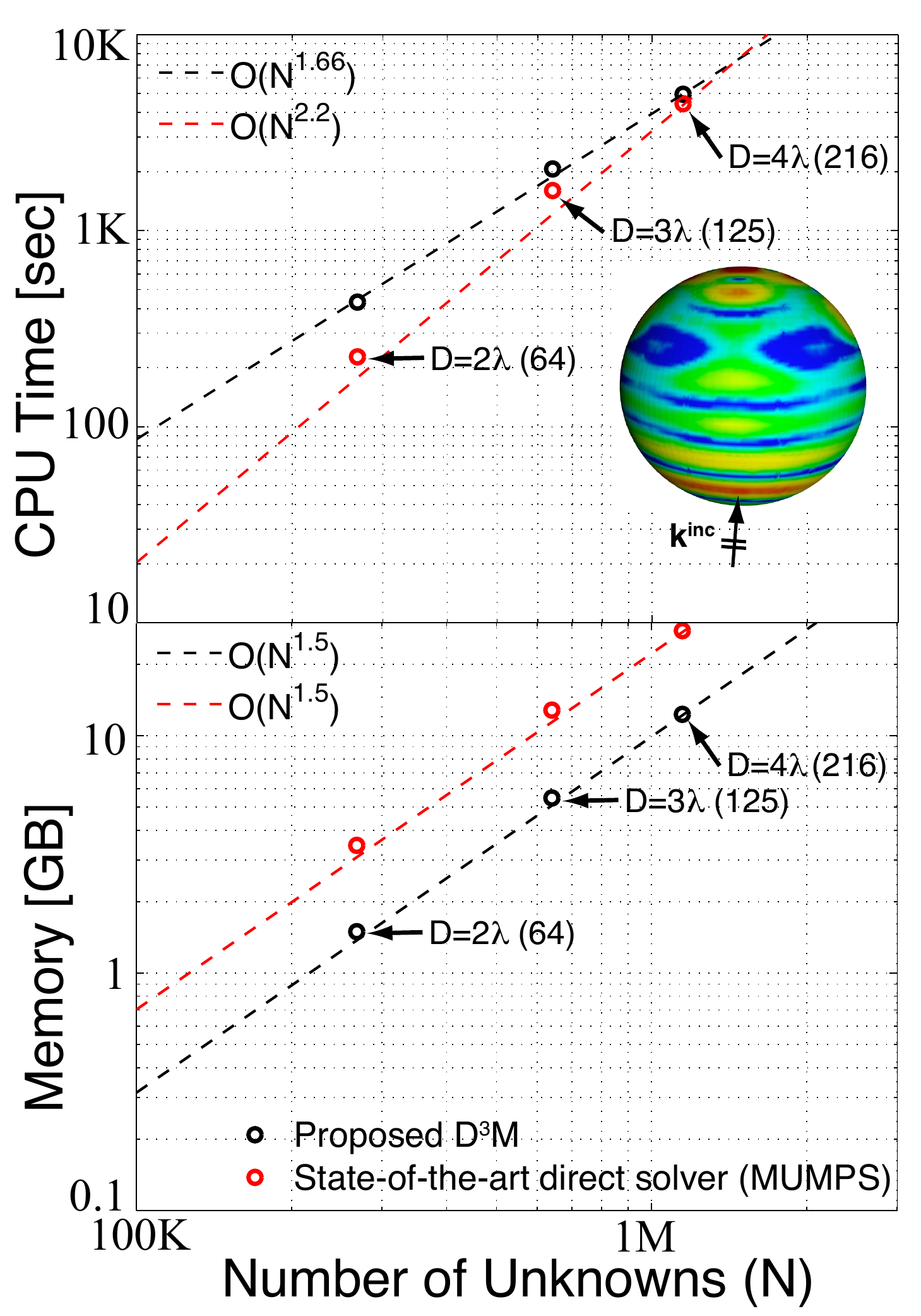}}
		\caption{Computational complexity for progressively larger dielectric spheres using the proposed D$^3$M and MUMPS: Factorization time (top); Factorization memory (bottom).}\label{fig:sphere}
	\end{center}
 \end{figure}

Next, the scattering of progressively larger dielectric spheres is considered. Computational complexity of factorization time and memory for these problems using the proposed D$^3$M and MUMPS are shown in Fig. \ref{fig:sphere}. Again, the proposed method uses more than 2.25 times less memory and surprisingly better time complexity than MUMPS. The relative residual error ($\|{\bf A} \bar{x} - f \|_{inf} / \|f\|_{inf}$) of all runs using the proposed D$^3$M is around $10^{-12}\!\!-\!\!10^{-13}$ which was the same as MUMPS.

\begin{thebibliography}{1}

\bibitem{IEEEhowto:moshfegh2019parallel}
J. Moshfegh,  D. G. Makris, and M. N. Vouvakis,``Parallel Direct Domain Decomposition Methods (D$^3$M) for Finite Elements." \emph{2019 IEEE International Symposium on Antennas and Propagation and USNC-URSI Radio Science Meeting}, pp. 777--778. \label{ref:moshfegh2019}

\bibitem{IEEEhowto:zhu2006}
Y. Zhu and A. C. Cangellaris, Multigrid finite element methods for electromagnetic field modeling. \emph{John Wiley \& Sons},  Vol. 28, 2006. \label{ref:zhu2006}

\bibitem{IEEEhowto:toselli2005}
A. Toselli and O. Widlund, \emph{Domain decomposition methods: algorithms and theory}, Vol. 3. Berlin: Springer, 2005. \label{ref:toselli2005}

\bibitem{IEEEhowto:amestoy2001}
P. R. Amestoy, I. S. Duff, J. Y. L'Excellent, and J. Koster, ``A fully asynchronous multifrontal solver using distributed dynamic scheduling," \emph{SIAM Journal on Matrix Analysis and Applications} Vol. 23, No. 1, pp.15-41, 2001. \label{ref:amestoy2001}

\bibitem{IEEEhowto:ying2014}
P. G. Schmitz, and L. Ying, ``A fast nested dissection solver for Cartesian 3D elliptic problems using hierarchical matrices," \emph{Journal of Computational Physics}  Vol. 258, pp.227-245, 2014. \label{ref:ying2014}



\bibitem{IEEEhowto:bunch1977}
J. R. Bunch, L. Kaufman, ``Some stable methods for calculating inertia and solving symmetric linear systems," \emph{Math. Comp.}, Vol. 31, No. 137, pp. 163–179, 1977.
\label{ref:bunch1977}


%
%





%
%



\end{thebibliography}
\end{document}